\documentclass[aps,prb,twocolumn,letterpaper]{revtex4}
\usepackage{graphicx}
\usepackage{dcolumn}
\usepackage{amsmath}
\usepackage{amssymb}
\usepackage{float}
\usepackage{epstopdf}
\usepackage{color}

\def\rv{{\bf r}}
\def\kv{{\bf k}}
\def\Sv{{\bf S}}
\def\Qv{{\bf Q}}
\newcommand{\beq}{\begin{equation}}
\newcommand{\eeq}{\end{equation}}
\def\beqa{\begin{eqnarray}}
\def\eeqa{\end{eqnarray}}

\usepackage{pdfpages}

\begin{document}

\title{Incommensurate spin density wave as a signature of spin-orbit coupling}
\author{Aaron Farrell$^1$, P.-K. Wu$^2$, Y.-J. Kao$^{2,3}$ and T. Pereg-Barnea$^1$}
\affiliation{$^1$Department of Physics and Center for the Physics of Materials, McGill University,
Montreal, QC, Canada\\
$^2$Department of Physics and Center of Theoretical Sciences, National Taiwan University, Taipei 10607, Taiwan\\
$^3$Center for Advanced Study in Theoretical Science, National Taiwan University, Taipei 10607, Taiwan}
\date{\today}
\begin{abstract}
Close to half filling and in the limit of strong interactions the Hubbard model leads a spin Hamiltonian.  In its original isotropic form this is the Heisenberg model with antiferromagnetic coupling $J$.  On a square lattice there is no frustration and the ground state is a classical antiferromagnet denoted by the commensurate order vector $\Qv=(\pi,\pi)$.
In this work we show that the inclusion of spin orbit coupling (SOC) in the fermionic Hubbard model on a square lattice leads to an incommensurate spin density wave (ISDW) in the strong interactions limit at half filling.
In this limit the Hubbard model leads to a non-diagonal and anisotropic  spin Hamiltonian which exhibits the incommensurate spin order as its classical ground state.  We note that an ISDW can be found in systems regardless of spin orbit-coupling due to nesting  and lattice distortion.  These effects, however, can usually be recognized by experimental probes.
\end{abstract}
\maketitle
Strongly correlated electron systems and reduced dimensionality are fertile grounds for novel physical phenomena.  It is in the strong interaction regime that deviations from Landau's Fermi liquid occur and quantum mechanics is manifest in macroscopic quantities.  Examples of such systems include the Fractional quantum Hall effect and high T$_c$ superconductivity.  On the other hand, spin orbit coupling is capable of producing non-trivial effects even when the electrons are essentially non-interacting.  In particular, it leads to the spin Hall effect\cite{Sinova}, the quantum spin Hall effect and other topological insulator states\cite{KaneMele1,KaneMele2} in two and three dimensions.  It is therefore tempting to study the combined effect of strong electron-electron interactions and spin orbit coupling.  This general question has been explored in different contexts giving rise to both topological behavior and non-trivial spin structures\cite{Cole,Radic,Minar,Witczak1,Witczak2}.  In bosonic systems\cite{Cole,Radic} the interest is motivated by cold atoms with synthetic gauge fields that give rise to Dzyaloshinskii-Moriya terms in the spin Hamiltonian while in fermionic systems\cite{Minar,Witczak1,Witczak2} the motivation is usually topology.

Two of us have previously studied the effect of strong correlations which lead to superconductivity in the presence of spin orbit coupling.  We found that the combined effect may, under certain circumstances, lead to topological superconductivity\cite{Farrell1,Farrell2}.  In this work, we take the strong interaction limit and specialize to the case of a half filled band.  This leads to a spin Hamiltonian in which the spin-spin interaction is non-diagonal and anisotropic.  This interaction contains Dzyaloshinskii-Moriya and compass anisotropy in addition to the usual
Heisenberg coupling. The combination of SOC and strong Coulomb repulsion leads to a unique spin ordered ground state (ISDW).  This ground state can be detected by magnetic probes such as neutron scattering and signal the significance of SOC.  While structural distortion and nesting may also lead to an ISDW, they can usually be seen in experimental probes. Once strong SOC is established topological behavior can be switched on through electron/hole doping.

In this Letter we study a spin Hamiltonian which is the strong interaction limit of the Hubbard model on a 2D square lattice with spin orbit coupling. The full derivation of this Hamiltonian can be found in Ref.~[\onlinecite{Farrell2}].  Here we simply sketch the main steps. We study this model by systematically searching for a spin configuration which minimizes the interaction energy. This is done in two ways: analytically by a variational ansatz spin-configurations and numerically via Monte Carlo (MC).  The two methods agree very well in most of our parameter space and we obtain a spin phase diagram, shown in Fig.~\ref{fig:PD1}. The phase diagram includes both commensurate and incommensurate spin arrangements with the incommensuration becoming more pronounced with increased spin-orbit coupling.

The extended Hubbard model we consider is given by the following Hamiltonian on a square lattice:
\begin{equation}
H=T+H_{SO}+H_U
\end{equation}
where
\begin{equation}
T = -\frac{t}{2}\sum_{i,\delta,\sigma} (c_{i,\sigma}^\dagger c_{i+\delta,\sigma}+ c_{i+\delta,\sigma}^\dagger
c_{i,\sigma})
\end{equation}
is the tight binding kinetic energy and
\begin{equation}
H_{SO} = \sum_{\kv} \psi_{\kv}^\dagger \mathcal{H}_\kv \psi_{\kv},
\end{equation}
is the spin-orbit coupling part of the Hamiltonian.  Here $\mathcal{H}_\kv = {\bf d}_\kv \cdot \vec{ \sigma}$ with ${\bf d}_\kv=(A\sin{k_x}, A\sin{k_y}, 2B(\cos{k_x}+\cos{k_y}-2)+M)$.  The constant $A$ is the usual Rashba spin-orbit coupling.  The constants $B$ and $M$ polarize the spin in the $z$ direction.  $M$ is the standard Zeeman field and $B$ appears in the 2D quantum well topological insulator model known as the BHZ model\cite{BHZ}.  While the Rashba term $A$ is responsible for the incommensurate spin density wave, the Zeeman like terms tilt the magnetization outside of the plane and eventually favor ferromagnetism.

 The interaction energy
\beq
H_U = U \sum_i n_{i\uparrow} n_{i,\downarrow}
\eeq
describes on-site repulsion.  This Hamiltonian, with additional off-site interaction, was studied in Refs.~\onlinecite{Farrell1,Farrell2}.  Here we are concerned with its strong interaction limit at half filling.  The full details of the strong coupling expansion at arbitrary filling can be found in Ref.~\onlinecite{Farrell2}. 

The strong coupling expansion distinguishes between the high interaction energy scale $U$ and other, quadratic terms in the Hamiltonian.  Written in real space the quadratic terms are either on-site (chemical potential) or hopping-like.  The on-site terms do not change the potential energy while the hopping terms do.  The hopping terms are either spin independent and spin conserving (the usual hopping) or spin dependent/spin flip hopping which are the real space version of $H_{SO}$.  Ideally, one would like to diagonalize the Hamiltonian but due to the presence of both quadratic and quartic terms this can not be done analytically.  Instead of diagonalizing we set out to block diagonalize. Since we are interested in the strong coupling regime we would like to block-diagonalize the Hamiltonian such that the interaction energy is constant at each block.  In other words - there exist a unitary transformation which eliminates terms which change the interaction energy.  Using the desired properties of the transformation (unitarity and interaction energy conservation) we can formally write down the transformation.  In order to find a closed form, however, we resort to a power expansion in a small parameter such as $1/U$.  We are then able to find the transformation and the transformed Hamiltonian up to a given order in $1/U$.

At half filling and in the limit of infinite $U$, the system is uniformly charged with one electron per lattice site. At finite $U$ we therefore expect the transformation to eliminate the hopping terms which change the uniform charge distribution to a state with doubly occupied sites.  The first non-zero term in the expansion of the transformed Hamiltonian is therefore a sum of all double hopping processes in which a state with a doubly occupied site is virtually created and annihilated.  In the isotropic Hubbard model this leads to the Heisenberg model with the coupling $J = 4t^2/U$, with $t$ stemming from the double hopping process.  In the present case the same considerations lead to:
\beq
{\cal H} = \sum_{i,\delta}\sum_{\alpha\beta=\{x,y,z\}}J_\delta^{\alpha\beta}S_i^\alpha S_{i+\delta}^\beta +\sum_i h_z S_i^z
\label{eq:Spin_Hamiltonian}
\eeq
where $i$ enumerates all lattice sites and $i+\delta$ are the nearest neighbors of $i$.  $\alpha$ and $\beta$ are spin directions, $h_z=2M-8B$ is the Zeeman field and the coupling matrix is given by:
\begin{eqnarray}
 &J_{\delta}={1\over 2 U}\times \\ \nonumber  &\left(\begin{matrix}
      4t^2+{A}^2a(\delta)-4B^2 & 0& -4At\delta_y \\
      0 & 4t^2-{A}^2a(\delta)-4B^2&4At\delta_x \\
      4At\delta_y&-4At\delta_x&4t^2-A^2+4B^2\\
   \end{matrix}\right)
\end{eqnarray}
where $a(\delta) =A(\delta_x-\delta_y)$.  We refer to this model as a non-diagonal and anisotropic Heisenberg model.  We note that the coupling matrix can not be diagonalized by a simple axis rotation due to its dependence on the nearest neighbor vector $\delta$.  It is also useful to write the Hamiltonian in momentum space where it takes the form:
%\beqa
%J(\kv)&=& J_1(\cos(k_x)+\cos(k_y)) + J_2(\cos(k_x)-\cos(k_y)) \nonumber \\ &+& J_3\sin(k_x)+J_4\sin(k_y)
%\eeqa
%where
%\begin{eqnarray}
% J_1= {1\over U}\left(\begin{matrix}
%      4t^2-4B^2 & 0& 0 \\
%      0 & 4t^2-4B^2&0 \\
%      0&0&4t^2-A^2+4B^2\\
%   \end{matrix}\right)
%\end{eqnarray}
%\begin{eqnarray}
%  J_2= {1\over U}\left(\begin{matrix}
%      A^2& 0& 0\\
%      0 &-A^2&0 \\
%     0&0&0\\
%   \end{matrix}\right)
%\end{eqnarray}
%\begin{eqnarray}
% J_3= {1\over U}\left(\begin{matrix}
%      0 & 0& 0 \\
%      0 & 0&4Ati \\
%      0&-4Ati&0\\
%   \end{matrix}\right)
%\end{eqnarray}
%\begin{eqnarray}
%  J_4= {1\over U}\left(\begin{matrix}
%     0& 0& -4Ati \\
%      0 & 0&0 \\
%      4Ati&0&0\\
%   \end{matrix}\right)
%\end{eqnarray}
%and the Hamiltonian takes the form:
\beq
{\cal H} = \sum_\kv J(\kv)^{\alpha\beta}\Sv^\alpha_\kv\Sv^\beta_{-\kv}+h_z\Sv_{\kv=0}
\eeq
where $J(\kv)$ is the Fourier transform of $J_\delta$.

Anisotropic Heisenberg Hamiltonians have been studied by the Luttinger-Tisza (LT) method\cite{LT1,LT2, Lyons} in the context of a dipolar crystal fields and spinel states in three dimensions.  We generalize this method for the non-diagonal anisotropic case in the presence of a Zeeman field but unfortunately, find that it fails to find a valid spin configuration.  The failure is manifested in spin configurations which are not properly normalized at each lattice site (the so-called ``detailed constraint"\cite{LT1,LT2, Lyons}). Notwithstanding, this method still gives us inspiration for a variational study in which the variational spin configuration we use is similar to the states obtained by the LT method.  We sketch the results of the LT method here for the sake of clarity and completeness. In this method the spins are represented by classical vectors at each lattice site and the problem of finding the ground state amounts to minimizing the energy function under the constraint that the vectors are of length $S$.  The constraint is enforced by introducing Lagrange multipliers. Since the spin vector is defined on $N$ sites this leads to $N$ Lagrange multipliers which results in a system of equations that can not be solved analytically.  Luttinger and Tisza suggested to impose the constraint only on the average spin and minimize the Lagrangian function:
\beq\label{eq:Lagrangian}
\mathcal{L} = H - \lambda \left(\sum_{i} \Sv^2(\rv_i)  - NS^2\right)
\eeq
While this, in general, can be done analytically, it might yield a solution that does not obey the detailed constraint.  In some cases\cite{Lyons} the solutions obtained in this way have enough freedom that a solution which satisfies the detailed constrained can be constructed.  In our non-diagonal model this is not the case.  Nevertheless insight into the problem is gained from this method.  In the absence of Zeeman field the minimization of Eq.~(\ref{eq:Lagrangian}) leads to the condition $[J(\kv)-\lambda]\Sv_\kv=0$.
%\beq\label{eq:lambda}
%[J(\kv)-\lambda]\Sv_\kv=0
%\eeq
This tells us that if $\Sv_\kv \neq 0$ then $\lambda$ must be an eigenvalue of the matrix $J(\kv)$.  Since $\lambda$ is not a function of $\kv$, $\Sv_\kv$ must be zero for all but a few values of $\kv$. Let us call one of these values $\Qv$, then owing to the time reversal invariance of $J(\kv)$ the second value of $\kv$ is $-\Qv$. This leads to a spin state $\Sv_\kv =\Sv_{\Qv}\delta_{\kv,\Qv}+ \Sv^*_{-\Qv}\delta_{\kv,-\Qv}$.  Furthermore, plugging the condition $[J(\kv)-\lambda]\Sv_\kv=0$ back into the Hamiltonian shows that in order to minimize the energy one has to minimize $\lambda$.  It is therefore clear that the ground state configuration will be found through diagonalizing the coupling $J(\kv)$ and minimizing $\lambda$ by choosing the appropriate order vector $\Qv$.  All other Fourier components of $\Sv$ will be set to zero. Note that due to symmetry there is usually more than one order vector but all $\Qv$ are related by the symmetry in the system.

Following the LT method in our system yields spin states of the forms $\Sv_{\Qv} = A(a,0,ib)^T$ and $\Sv_{\Qv} = A(0,a,ib)^T$ with an incommensurate $\Qv$.  While these minimize the energy they do not allow enough freedom for normalizing the spin vector on each site.  Inspired by the LT method we look for solutions to the classical Hamiltonian which posses only a few ordering vectors $\Qv$, minimize the energy {\it and} obey the detailed constraint.  We do so by proposing a few different ans\"atze with the order vector $\Qv$ as one of the free parameters.  Each ansatz is similar to the eigenstate found in the LT method, however it is built to obey the normalization constraint.

The first ansatz describes an incommensurate spin density wave and is a normalized version of the LT state given by
\begin{eqnarray}
&\Sv_{ISDW}(\rv_i)= \\ \nonumber &\left(\cos(\phi)\cos(\Qv\cdot \rv_i),\sin(\phi)\cos(\Qv\cdot \rv_i), -\sin(\Qv\cdot \rv_i)\right)^T
\end{eqnarray}
where we work in units where $S=1$. By construction the above ansatz automatically satisfies $\Sv(\rv_i)^2=1$ and we use it as a variational state. The variational energy reads
\beqa
E_{ISDW}(\Qv,\phi) &=& \left(4t^2-\frac{A^2}{2}\right)\epsilon_{\Qv}+ \frac{1}{2}\cos(2\phi)A^2\bar{\epsilon}_{\Qv} \nonumber \\ &+& 4At(\sin(\phi)  \xi_{Q_x}-\cos(\phi)\xi_{Q_y})
\eeqa
where $\epsilon_{\Qv}= {1\over U}(\cos(Q_x)+\cos(Q_y))$, $\xi_{Q_i} = {1\over U}\sin(Q_i)$ and $\bar{\epsilon}_{\Qv}= {1\over U}(\cos(Q_x)-\cos(Q_y))$.

Minimizing with respect to $Q_x, Q_y$ gives
\beqa\label{ks}
Q_x &=& n_x\pi - \arctan\left[\frac{4At\sin{\phi}}{A^2\sin^2{\phi}-4t^2}\right] \\ \nonumber
Q_y &=& n_y\pi -\arctan\left[\frac{4At\cos{\phi}}{A^2\cos^2{\phi}-4t^2}\right]
\eeqa
where the branch in each case ({\it i.e.} the value of the integers $n_x,n_y$) is chosen such that $
\text{sgn}(\cos{Q_i})=-\text{sgn}\left(1-\left(\frac{A^2(1+g_i\cos(2\phi))}{8t^2}\right)^2\right)$ where $g_x=-g_y=-1$.
%\beqa\label{ks}
%\text{sgn}(\cos{Q_x}) &=&-\text{sgn}\left(1-\left(\frac{A^2(1-\cos(2\phi))}{8t^2}\right)^2\right) \\ \nonumber
%\text{sgn}(\cos{Q_y}) &=&-\text{sgn}\left(1-\left(\frac{A^2(1+\cos(2\phi))}{8t^2}\right)^2\right)
%\eeqa
For our parameter range $|A|<2t$ and $n_x=1$ and $n_y=1$. Plugging the above expressions into the energy we find the minimum energy:
\beq
E_{ISDW}^{\text{Min}} = -{1\over U}\left(A^2+8t^2\right)
\eeq
which is independent of $\phi$.  This can be thought of as an easy axis ISDW. 
%%% Added by YJKAO
{Furthermore, we perform classical  MC simulation on Eq.~(\ref{eq:Spin_Hamiltonian}) to find the ground state spin configuration.
 We use  a parallel Metropolis single-spin  update\cite{GPU} and over-relaxation method\cite{OR} on a graphic processing unit (GPU). 
In addition, we use the parallel tempering algorithm\cite{PT}  as an optimization method to find the ground states for a given parameter set.   In the simulation, one Monte Carlo step(MCS) involves one single-spin update, one over-relaxation move and one parallel tempering swap. In  a typical simulation,  10$^6$ MCS's are performed to ensure the system is in equilibration and the lowest temperature configuration is taken as the ground state configuration for the given parameter set.\cite{Supp} }
%%%%
From the  simulation, we find a spin wave in some arbitrary direction.  In order to compare the result to our ansatz, we simply read off $\phi$ from the simulation.  With a given $\phi$ we calculate $Q_{ISDW} = \sqrt{Q_x^2+Q_y^2}$ and compare it with the periodicity of the MC data.  The result for different values of the spin-orbit coupling $A$ are shown in Fig.~\ref{fig:comparison}. A simple expression for the dependence of $Q_{ISDW}$ on $A$ can be obtained by taking $\phi =0$ which gives $Q_y=\pi$ and $Q_x=Q^x_{ISDW} = \pi-\arctan\left[\frac{4At}{4t^2-A^2}\right]$

\begin{figure}[tb]
  \setlength{\unitlength}{1mm}

   \includegraphics[scale=.36]{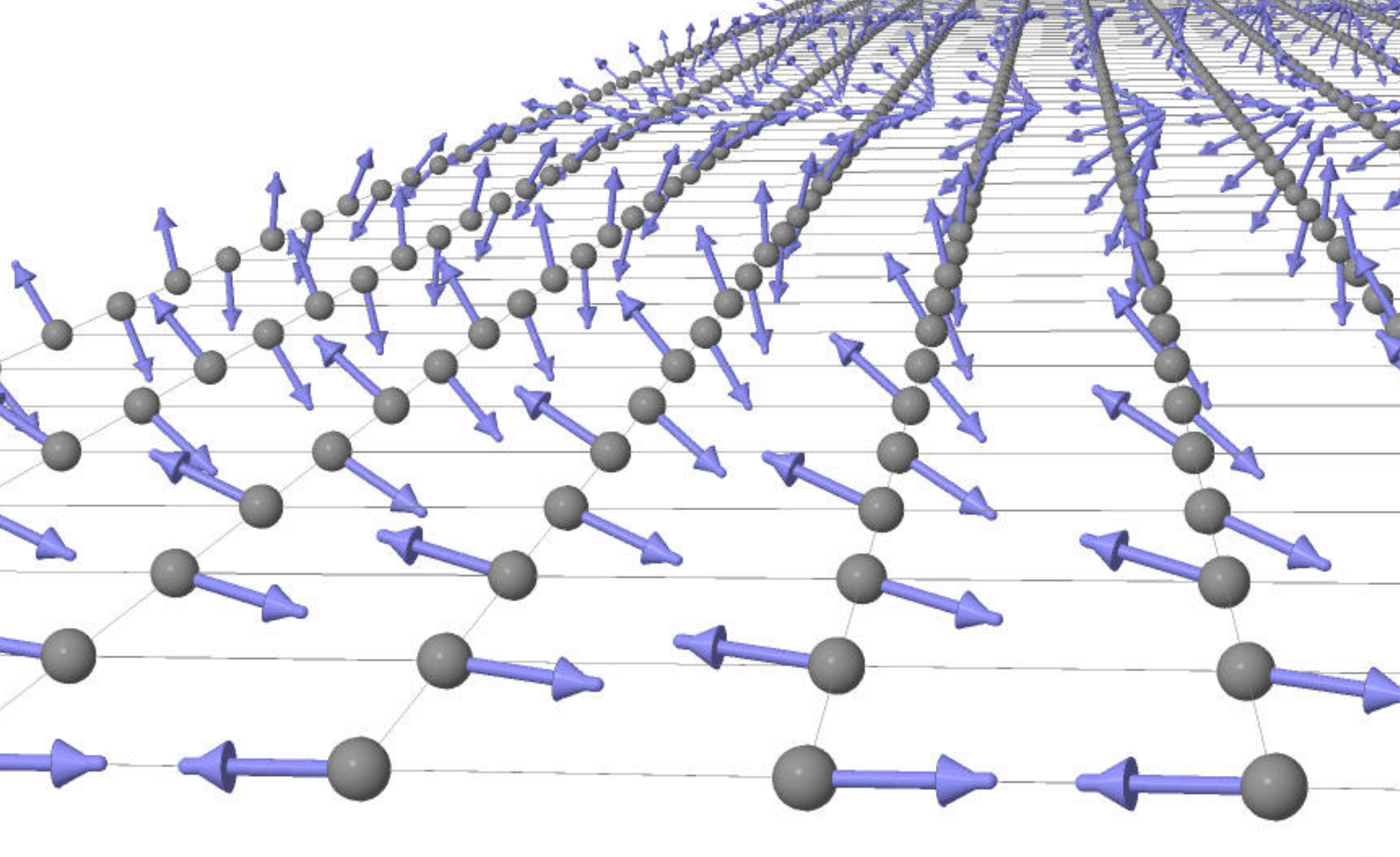}
\caption{{\small
Visualization of spin configuration in state A. The figure shows the spin on each lattice site as seen by looking along the y-axis.
%Visualization of spin configuration in state A. The main figure shows the spin on each lattice site as seen by looking along the y-axis. The inset in the figure shows a plot of the minimizing function $Q^x_{ISDW}$ (as defined in the text) as a function of the spin-orbit coupling strength $A$ scaled in units of $t$. For ease of comparison with the point $(\pi,\pi)$ we have scaled $Q^x_{ISDW}$ by $\pi$. In the main figures we have taken $A=0.15t$ which corresponds to $Q_x \simeq  2.9919$.
     }
     }\label{fig:KA}
\end{figure}

For small $A$ our wave vector is close to $\pi,\pi$ and it is therefore constructive to think of the spins as flipping their direction between nearest neighbors with additional spin rotation on longer length scales.  This is illustrated in Fig.~\ref{fig:KA}.  We emphasize that the wave vector of the spin modulation need not be a rational fraction of $\pi$.

We now account for a possible Zeeman field in the form of $M,B\ne0$.  In the LT method this changes the relation between the eigenvalue $\lambda$ and the energy, however, it leaves it monotonically increasing in the relevant range such that the procedure remains the same.  We therefore add to our set of variational solutions two more general configurations:
\beq
\Sv_{TAF}(\rv_i)= \left(\sin(\phi)\cos(\Qv\cdot \rv_i),\sin(\phi)\sin(\Qv\cdot \rv_i), \cos(\phi)\right)^T,
\eeq
where $\Qv$ and $\phi$ are to be optimized and
\beq
\Sv_{FM}(\rv_i)= \left(0,0,\text{sgn}(4B-M) \right)^T.
\eeq

The second state $\Sv_{TAF}$ is a generalization of the previous ansatz in the presence of a Zeeman field.  We label it by TAF to denote a tilted antiferromagnet as will be made clear shortly.  The third state is simply a ferromagnetic state which is expected at high Zeeman fields.

Minimization of the energy of the tilted antiferromagnet can done analytically; its variational energy is given by\\
\beqa
E_{TAF} &=& \frac{\cos^2(\phi)(8t^2-2A^2+8B^2)}{U}+(2M-8B)\cos(\phi) \nonumber \\ &+&\frac{4\sin^2(\phi)(t^2-B^2)(\cos(Q_x)+\cos(Q_y))}{U}
\eeqa
Inspecting the above we find that  (assuming $|t|>|B|$) we must pick $\Qv=(\pi,\pi)$ and $\phi = \arccos\left[\frac{Uh_z}{4\left(A^2-8t^2\right)}\right]$
%\beq
%\phi = \arccos\left[\frac{U(M-4B)}{2\left(A^2-8t^2\right)}\right]
%\eeq
which is valid in the region where $U|h_z|<4\left(|A^2-8t^2|\right)$.
Plugging this solution into the energy gives:
\beq
E_{TAF}^{Min} = {1\over U} \left(8B^2-8t^2\right) +\frac{U(4B-M)^2 }{2\left(A^2-8t^2\right)}
\eeq
In addition we calculate the energy of the ferromagnetic state, $\Sv_{FM}$.  This gives:
\beq
E_{FM} = -2|M-4B| +{2\over U}(4t^2-A^2+4B^2)
\eeq

\begin{figure}[t]
  \setlength{\unitlength}{1mm}

   \includegraphics[scale=.35]{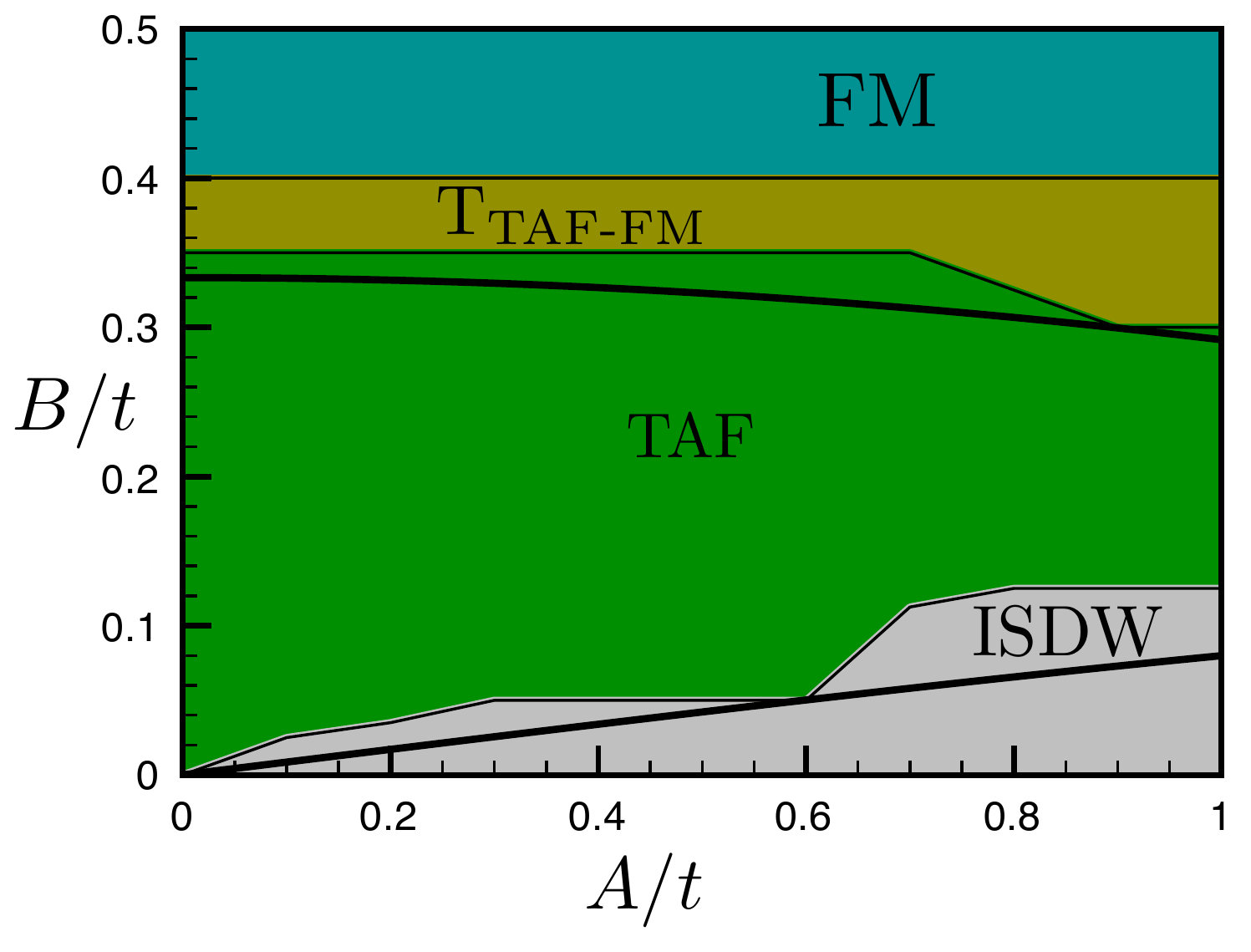}
\caption{{\small
Phase diagram of our model using both the {\em ansatz} method and MC data.  We have set $M=0$ and $U=12t$. The different colors represent phases in the numerical MC data which are very similar to the phases found in our VM theory. The region we have labelled ISDW is qualitatively similar to our ISDW phase, the TAF phase is similar to the state $\Sv_{TAF}$ and FM labels a state very similar to our $\Sv_c$. We have also used a label $ \text{T}_{\text{TAF-FM}}$ which is a transition region between a TAF and a FM. For reference we have included dark lines in this figure to show the phase boundaries as determined by our variational calculations. The crude boundaries between different MC phases are a result of the numerical limitation of having data for only a finite number of lattice points.
%We have labeled the $\Sv_{ISDW}$ state with an ``ISDW", the $\Sv_{TAF}$ with a ``TAF" and the $\Sv_c$ with a ``FM".
     }
     }\label{fig:PD1}
\end{figure}

We use the three types of ansatz above to calculate the energy as a function of different parameters.  A phase diagram for our ansatz method as well as for numerical MC results is presented in Fig.~\ref{fig:PD1}.  The ISDW phase occurs due to spin orbit coupling, as can be seen from Eq.~\ref{ks}.  Its ordering vector for the particular choice of $\phi=0$ is $\Qv=(\pi,K(A))$ where $K(A=0)=\pi$ (when there is no SOC) and decreases monotonically with increasing SOC parameter $A$.  This phase is the ground state for any non-zero SOC when there is no Zeeman field.  When the Zeeman field is turned on the spins start to tilt into the field direction.  In this state the spin projection on the $x$-$y$ plane is antiferromagnetically ordered and no incommensuration is detected, we refer to this as the tilted antiferromagnet (TAF).  The boundary between the ISDW phase and the TAF occurs when the Rashba strength reaches the following critical value $A_{c1} = \sqrt{4t^2-4B^2-\sqrt{(4B^2+4t^2)^2-{U^2\over 8}h_z^2}}$
%\beq
%A_{c1} = \sqrt{4t^2-4B^2-\sqrt{(4B^2+4t^2)^2-{U^2\over 2}(M-4B)^2}},
%\eeq
and the boundary between the TAF state and the ferromagnetic state is given by $|A_{c2}| = \sqrt{8t^2-\frac{U}{4}|h_z|}$
%\beq
%|A_{c2}| = \sqrt{8t^2-\frac{U}{2}|M-4B|},
%\eeq
such that below $A_{c1}$ we have the ISDW phase, between $A_{c1}$ and $A_{c2}$ the TAF phase and above $A_{c2}$ the ferromagnetic phase.

\begin{figure}[tb]
  \setlength{\unitlength}{1mm}

   \includegraphics[scale=.42]{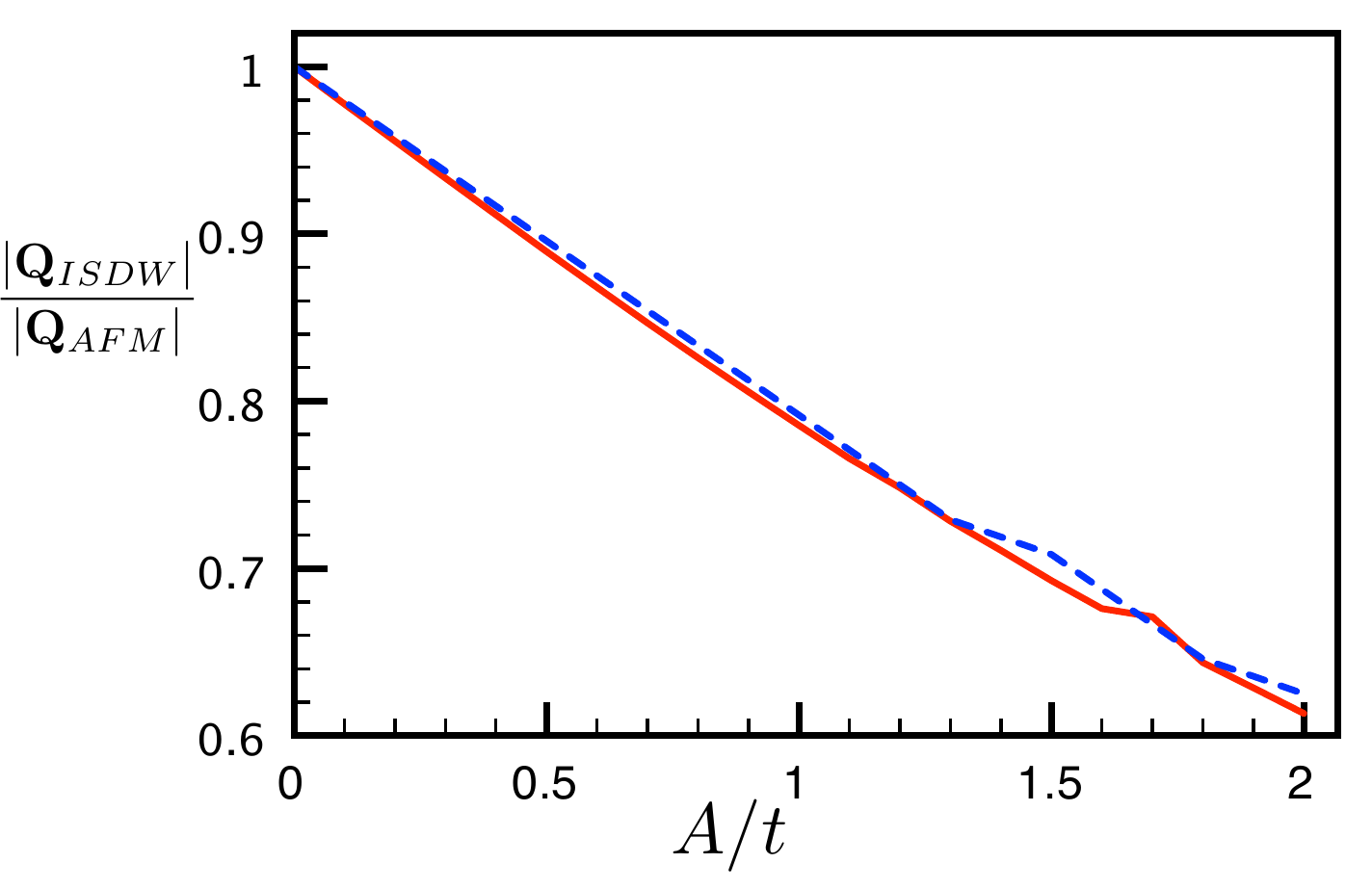}
\caption{{\small
Comparison of $|\Qv_{ISDW}|$ as found from VM and MC calculations. In the figure the solid (red online) curve is obtained from our VM theory while the dashed curve (blue online) is obtained from MC data. We have scaled the vertical axis by the magnitude of $\Qv_{AFM}=(\pi,\pi)$.
     }
     }\label{fig:comparison}
\end{figure}

To close we briefly provide the highlights of the comparison between our variational data and numerical MC data, leaving additional details to [\onlinecite{Supp}]. The agreement between MC data and the analytic method formulated here is extremely good, especially considering the simplicity of our approach. Where our method predicts an ISDW, a TAF or a FM the MC data gives spin configurations very consistent with these phases. A quantitative comparison of how close this agreement is, is shown in Fig.~\ref{fig:comparison} where we plot the $\Qv$ vector for the ISDW for various values of $A$. Slight disagreements appear in the range $.05t<B<.15t$ and $A>0.7t$, where variational theory predicts an TAF while the Monte Carlo data is more consistent with an ISDW. Another small disagreement between the two is at large $B$ where the transition between TAF and FM in the MC data is very gradual the variational data predicts a sharp transition.

In summary, we have shown that a spin Hamiltonian which represents the strongly interacting, half-filled limit of the Hubbard Hamiltonian with spin orbit coupling on a square lattice gives rise to an incommensurate spin density wave.  The ISDW appears only when Rashba type SOC is present and disappears if a strong enough magnetic field is applied.  This suggests that the identification of the ISDW in an isotropic system close to half filling may indicate the presence of significant spin orbit coupling.

The authors acknowledge useful discussions with M.Randaria and support from NSERC and FQRNT (TPB), the Vanier Canada Graduate Scholarship (AF), the Schulich Graduate Fellowship (AF), the Walter C. Sumner Memorial Fellowship (AF) and the Tomlinson Master's Fellowship (AF), and NSC in Taiwan through Grants No.~100-2112-M-002-013-MY3, 100-2923-M-004-002 -MY3, 102-2112-M-002-003-MY3 (PKW, YJK). TPB and YJK acknowledge the hospitality of the Aspen center for physics where some of this work has been carried out.

\bibliographystyle{apsrev}
\bibliography{SO-Heisenberg}



\section*{Supplementary material}

# Supplemental Material for Incommensurate spin density wave as a signature of spin-orbit coupling

## I. DETAILS OF MONTE CARLO CALCULATIONS

The variational method we have employed in the main text of this letter uses three highly simplified *ansätze* in order to gain some intuition of the underlying physics in our spin Hamiltonian. In order to evaluate how well these *ansätze* do we have carried out Monte Carlo simulations of the system described by our model spin Hamiltonian. To compare with our zero-temperature variational method approach we have run these simulations at very low (essentially zero) temperatures. The majority of our data was produced by running simulations on a $32 \times 32$ square lattice. In cases where we were skeptical that the data was effected by a finite size effects we opted to increase the lattice size to $96 \times 96$.

## II. QUALITATIVE COMPARISON

One can form a qualitative comparison between our analytic variational method (VM) and numerical MC results by simply plotting the real-space configuration of spin vectors given by each method. Several representative plots of these spins for two different phases for each the MC and the variational method are shown in Fig. S1. Depending on where in the phase diagram we are, the VM gives three qualitatively different configurations. Starting with the simplest, the FM phase shows all spins on each lattice site fully aligned with either the positive or negative z axis (depending on the sign of $4B-M$). The TAF phase has a few different characteristics. The spins on each site have a large and spatially constant value of $\mathbf{S}^z(\mathbf{r}_i)$ while the $xy$ projection of $\mathbf{S}(\mathbf{r}_i)$ in this phase alternates antiferromagnetically from one lattice site to the other with a wave vector $\mathbf{Q}_{TAF} = (\pi, \pi)$. All directions about which it can rotate are degenerate. Finally we have the ISDW phase which is spatially the most complicated. This phase showcases a spin state that propagates with a wave vector $\mathbf{Q}_{ISDW}$ in both the $z$ direction and in the $xy$ plane. In general we have $\mathbf{Q}_{ISDW} \neq (\pi, \pi)$ and $\mathbf{Q}_{ISDW}$ is weakly dependent on $\phi$, the angle about which the $xy$ projection of $\mathbf{S}_{ISDW}(\mathbf{r}_i)$ propagates. The energy of $\mathbf{S}_{ISDW}(\mathbf{r}_i)$ is degenerate in $\phi$. When this phase is viewed in real space the most notable characteristic is that stripes of large and small values of $|\mathbf{S}^z_{ISDW}(\mathbf{r}_i)|$ are clearly visible, the direction of these stripes depends on what choice of $\phi$ is made.

Depending on what part of the phase diagram we concern ourselves with, our VM predicts the ground state to be either an ISDW, a TAF, or a FM. Based on this prediction, we look for the characteristics associated with each phase (as outlined above) in the Monte Carlo data.

We find very good qualitative agreement between VM and MC data. Here we will highlight the comparison between the two sets of data in each phase. First, for $B = 0$ and $A > 0$ the MC data gives a spin configuration consistent with an ISDW. The wave vector of this ISDW in the MC data is slightly different than that of our VM calculation, but the agreement between the two is reasonable (more on quantitative comparisons in the next subsection).

Moving on to where VM predicts a TAF, in this region of the phase diagram we see a MC phase very similar to a $TAF$. This phase has a large $z$ component of $\mathbf{S}$ and an $xy$ projection whose direction oscillates in space. Deviations between our variational theory and MC calculations are that the value of $\mathbf{S}^z(\mathbf{r}_i)$ sometimes fluctuates very slightly in space and that the wave vector of the oscillations exhibited by the $xy$ projection of $\mathbf{S}(\mathbf{r}_i)$ is sometimes slightly different from the wave vector of $(\pi, \pi)$ predicted by the VM. There is one region of $AB$ space where there is stark *disagreement* between MC and VM data. This region is for $B = 0.1t$ and $A \geq 0.7t$. In this region the VM predicts a TAF while the MC data shows a phase consistent with an ISDW. It therefore appears that for larger values of $A$ the ISDW phase persists for larger values of $B$ than our VM predicts.

Finally, we focus on the FM phase. For the largest value of $B$ in our MC simulations we see a phase that is very close to a FM but not quite there. The value of $\mathbf{S}^z(\mathbf{r}_i)$ here lies in the region $0.95 \leq \mathbf{S}^z(\mathbf{r}_i) < 1$ but is not quite 1. Moreover there are still fluctuations of the $xy$ projection of $\mathbf{S}(\mathbf{r})$ and so this phase is still a TAF. Therefore in the MC data the phase boundary between the TAF and the FM phase occurs at larger values of $B$ than predicted by the VM. Also, the transition between these two phases is not as sudden as that between the ISDW and the TAF as the TAF can smoothly transition into a FM.

Using the above observations we have constructed a MC phase diagram for the model. This phase diagram along with plots of the VM phase boundaries appears in Fig. 1 of the main text.

## III. QUANTITATIVE COMPARISON

We now put the above qualitative observations onto more quantitative grounds by comparing some numerical features of the characteristics described above. The first characteristic we focus on is the wave vector $\mathbf{Q}_{ISDW}$ in the ISDW phase. We will compare the VM prediction for $\mathbf{Q}_{ISDW}$ to one obtained from MC data for different values of $A$ with $B$ fixed to zero. There are two technical details to address here before discussing results. First,

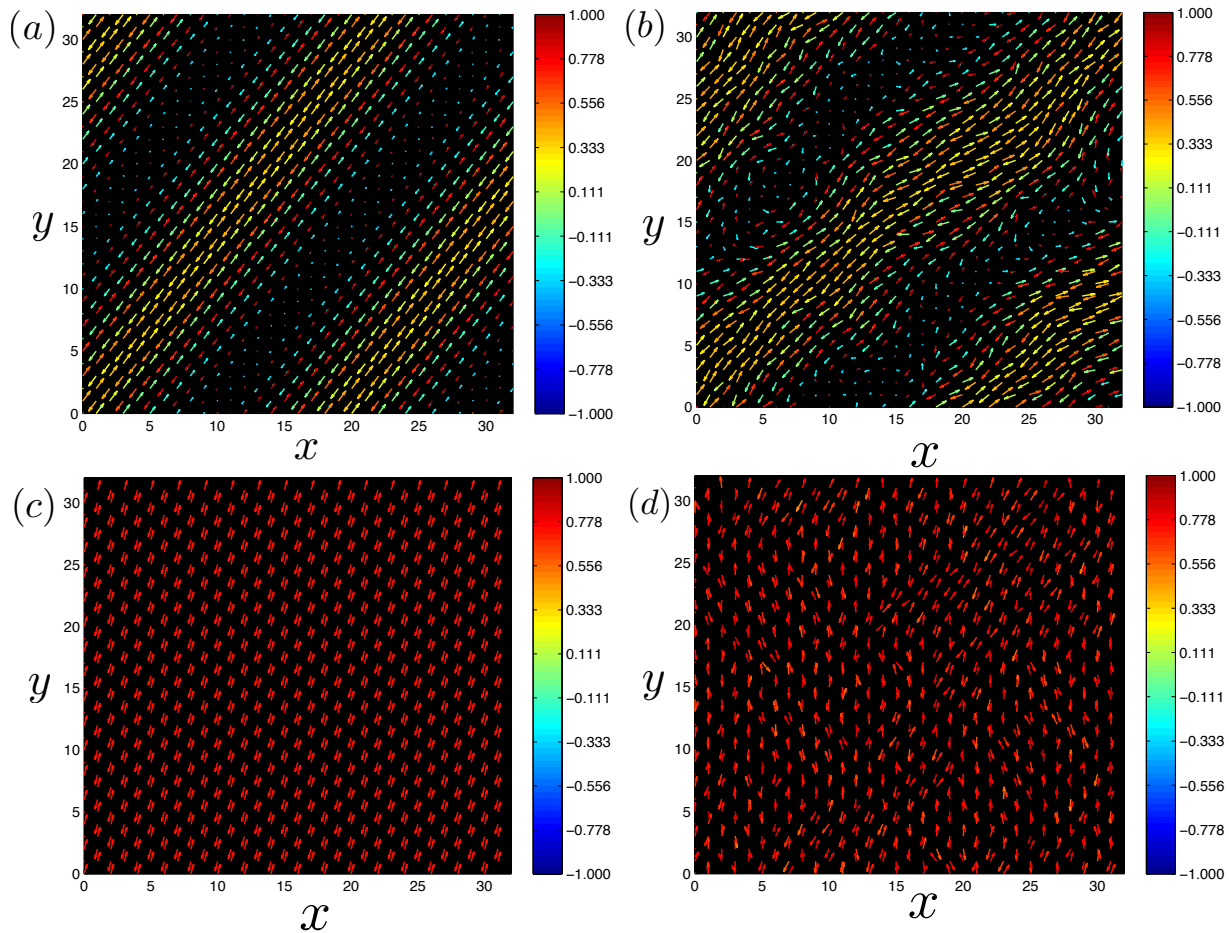

FIG. S1: Plot of spins in real space on a $32 \times 32$ lattice. In each case the arrows show the $xy$ projection of the spins while the colour map gives the $z$ values. The top line displays configurations for the variational (left) and MC (right) spends with $A = 0.2t$ and $B = 0$ (i.e. the ISDW phase) while the bottom line displays configurations for the variational (left) and MC (right) spends with $A = 0.1t$ and $B = .2t$ (i.e. the TAF phase).

the $\mathbf{Q}_{ISDW}$ result for the VM depends on the degenerate parameter $\phi$. In order to fix a value for $\phi$ we average the value of $\arctan\left[\frac{\mathbf{S}^y_{ISDW}}{\mathbf{S}^x_{ISDW}}\right]$ from MC data over each lattice site. This tells us what degenerate value of $\phi$ is most consistent with the particular MC configuration. Once we have done this we immediately have a VM prediction for $\mathbf{Q}_{ISDW}$. Second, in order to assign a value of $\mathbf{Q}_{ISDW}$ to the numerical data we form a Fourier decomposition as follows

$$\mathbf{S}_\mathbf{k} = \frac{1}{N} \sum_i e^{i\mathbf{k}\cdot\mathbf{r}_i} \mathbf{S}(\mathbf{r}_i) \qquad (1)$$

Inspecting the data we see quite generally that for $B$ fixed to zero (i.e. the ISDW phase) $|S^z_\mathbf{k}|$ is very strongly peaked about two vectors $\pm \mathbf{k}_{max}$. We then assign a MC value to $\mathbf{Q}_{ISDW}$ via $\mathbf{Q}_{ISDW} = \mathbf{k}_{max}$. To compare the $\mathbf{Q}$ vectors obtained from our analytic and numerical techniques we plot $|\mathbf{Q}_{ISDW}|$ as a function of $A$ for $B = 0$. This plot appears in Fig. 2 of the main text. Looking at Fig. 2 we see excellent agreement between the value of $\mathbf{Q}_{ISDW}$ obtained from the two methods. Not only does the VM method do a good job of reproducing the general trend of $|\mathbf{Q}_{ISDW}|$ as a function of $A$, but we also see good quantitative agreement between the two methods.

As one final comparison we investigate the in-plane wave vector of the MC data in regions of the phase diagram with $B > 0$. Based on our VM calculation we expect the ground state for $.1 \leq B \leq .3$ to be a TAF and as such we expect an in-place wave vector of $(\pi, \pi)$. To compare our predicted $\mathbf{Q}$ with that from MC we again compute a Fourier decomposition of $\mathbf{S}(\mathbf{r}_i)$ as shown in Eq. (1). Once we have this decomposition we look for the vector $\mathbf{k} = \mathbf{k}_{max}$ with the largest component of $|S^x_\mathbf{k}|$. To be consistent with our VM calculation this value of $\mathbf{k}$ should be $(\pi, \pi)$. We have plotted $|\mathbf{Q}| = |\mathbf{k}_{max}|$ as a function of $A$ for a few value of $B$ in Fig. S2.

Looking at Fig. S2 for $A \leq 0.7t$ we note that the agreement between VM and MC is very good as we see the dominant $\mathbf{Q}$ value for the MC data is exactly $(\pi, \pi)$. The exception to this is of course the $B = 0.1t$ curve for $A \geq .7t$ where, as we mentioned earlier, the MC data

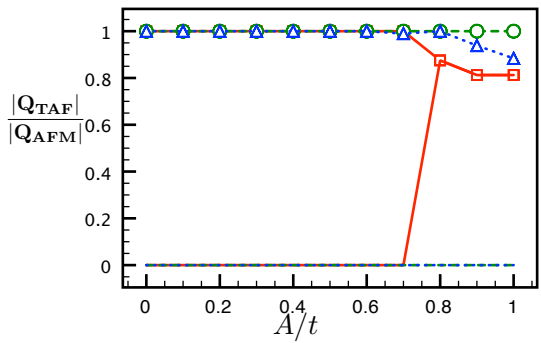

FIG. S2: Plot of $|\mathbf{Q}_{TAF}|$ as found MC data. In the figure the solid (red online) curves are for $B = 0.1t$, the dashed curves (green online) are for $B = 0.2t$ and the dotted curves (blue online) are for $B = 0.3t$. We have scaled the vertical axis by the magnitude of $\mathbf{Q}_{AFM} = (\pi, \pi)$. Data with symbols is for the $\mathbf{Q}$ vector in the $xy$ plane while data without symbols is for the wave vector of the $z$ projection.

gives an ISDW instead of a TAF. This is evident in the deviation of the $\mathbf{Q}$ vector of $S_x$ from $(\pi, \pi)$ and the sudden jump in the $\mathbf{Q}$ vector of $Sz$ from $(0, 0)$ seen in the figure. This part of Fig. S2 then signals the phase transition of the Monte Carlo data from a TAF to a ISDW as the dominant $\mathbf{Q}$ vector suddenly changes from $(\pi, \pi)$ to something incommensurate with the lattice. Another slight deviation occurs in the $B = 0.3t$ curve for large values of $A$, however the stark difference here is that the $\mathbf{Q}$ vector of $S_z$ remains zero here whereas for $B = 0.1t$ curve for $A \geq .7t$ it becomes finite. Although VM theory predicts the area of the phase diagram to be a TAF, we contend that this phase is actually in a transition phase between a TAF and a FM and is very close to the FM phase. This being the case, the $xy$ projection of $\mathbf{S}(\mathbf{r}_i)$ here is very small and the deviations in Fig. S2 are numerical fluctuations of the direction of this very small projection.

One should note that Fig. S2 can give a slightly deceiving picture of how close the MC phase is to our VM TAF phase. The dominant Fourier mode of the MC data in this part of the phase diagram is in excellent agreement with the TAF predicted by variational theory, however in some regions of $AB$ space there are weak sub-dominant Fourier modes in the MC $\mathbf{S}(\mathbf{r}_i)$. These sub-dominant modes do not show up in Fig. S2 but to contribute to slight fluctuations of the MC configuration away from the VM configuration. This subdominant behaviour is evident in our qualitative picture of the two spin results presented in Fig. S1. As mentioned early, in Figs. S1a and S1b we can see the "stripes" of the numerical ISDW phase have a different wavelength then our numerical results. Second, Figs. S1c and S1d shows a representative plot in the TAF phase where we see that the $xy$ projection of the spin can "drift" away from the $(\pi, \pi)$ oscillations predicted by our variational theory.



\end{document}